**Roadmap**

# 2022 Roadmap for Materials for Quantum Technologies

**Christoph Becher[1,*], Weibo Gao[2,*], Swastik Kar[3,*], Christian Marciniak[4], Thomas Monz[4,5], John G. Bartholomew[6], Philippe Goldner[7], Huanqian Loh[8], Elizabeth Marcellina[9], Kuan Eng Johnson Goh [8,10], Teck Seng Koh[9], Bent Weber[9], Zhao Mu[2], Jeng-Yuan Tsai[11], Qimin Yan[11], Samuel Gyger[12], Stephan Steinhauer[12], Val Zwiller[12]**

[1] Fachrichtung Physik, Universität des Saarlandes, 66123 Saarbrücken, Germany
[2] Division of Physics and Applied Physics, Nanyang Technological University, Singapore 637371
[3] Department of Physics, Northeastern University, Boston, MA 02115
[4] Institut für Experimentalphysik, 6020 Innsbruck, Austria
[5] AQT, 6020 Innsbruck, Austria
[6] Centre for Engineered Quantum Systems, School of Physics & Sydney Nanoscience Institute, The University of Sydney, Sydney, Australia
[7] Chimie ParisTech, CNRS, PSL University, Institut de Recherche de Chimie Paris, Paris, France
[8] National University of Singapore, Singapore
[9] Nanyang Technological University
[10] Agency for Science, Technology, and Research (A*STAR)
[11] Department of Physics, Temple University, Philadelphia, PA 19122, USA
[12] KTH Royal Institute of Technology, Stockholm 106 91, Sweden

[*] Editors of the Roadmap, to whom any correspondence should be addressed.
E-mails: <<christoph.becher@physik.uni-saarland.de; wbgao@ntu.edu.sg; s.kar@northeastern.edu>>

## Abstract

Quantum technologies are poised to move the foundational principles of quantum physics to the forefront of applications. This roadmap identifies some of the key challenges and provides insights on materials innovations underlying a range of exciting quantum technology frontiers. Over the past decades, hardware platforms enabling different quantum technologies have reached varying levels of maturity. This has allowed for first proof-of-principle demonstrations of quantum supremacy, for example quantum computers surpassing their classical counterparts, quantum communication with reliable security guaranteed by laws of quantum mechanics, and quantum sensors uniting the advantages of high sensitivity, high spatial resolution, and small footprints.  In all cases, however, advancing these technologies to the next level of applications in relevant environments requires further development and innovations in the underlying materials. From a wealth of hardware platforms, we select representative and promising material systems in currently investigated quantum technologies. These include both the inherent quantum bit systems as well as materials playing supportive or enabling roles, and cover trapped ions, neutral atom arrays, rare earth ion systems, donors in silicon, color centers and defects in wide-band gap materials, two-dimensional materials and superconducting materials for single-photon detectors. Advancing these materials frontiers will require innovations from a diverse community of scientific expertise, and hence this roadmap will be of interest to a broad spectrum of disciplines.

## Contents

# 1. Introduction

*Christoph Becher,[1] Weibo Gao,[2] Swastik Kar[3]*

The 21st Century is the age of quantum technologies, where the foundational principles of quantum physics such as coherence, superposition and entanglement of quantum states are tapped – often in devices as small as single defects and impurities – for trillion-time faster computations, beyond-classical ultrasensitive sensors, physically secure communications, and much more. While the scientific community and industry have made giant strides in some of these technologies, others are at earlier stages of technology readiness. In all cases, advancing these technologies to the next level will benefit substantially from materials advances and innovations. The goal of this roadmap is to identify some of these key challenges and provide insights on materials innovations required to "move the needle" further for a range of exciting quantum technology frontiers. In doing so, while we touch upon some relevant recent advances, this roadmap is not meant to be a review article or a progress report.

Innovations in quantum technologies traditionally (and largely) reside within the domain of physics and electrical engineering, with primary applications in the field of information science and technologies. However, we hope that this roadmap will appeal to a wider range of disciplines that encompass advancing materials science and engineering. Upstream in the innovation pipeline, materials scientists will play important roles in developing new or upgrading current methods for precision synthesis, scalable yield, and stable performance. At various levels of technology development, new tools and methods will be required for difficult-to-diagnose materials problems and reveal new opportunities. New methods and protocols might be needed to reduce noise and improve reliability. Developed technologies, while addressing traditional information science and related applications, could also potentially advance chemical, biological and other physical sciences. At each point, new advances in data mining and machine learning approaches will benefit materials science. Overall, this roadmap will provide a guideline to multidisciplinary research groups and diverse funding agencies the crucial advances needed, where the specific needs are, and how best to approach and address these needs.

In our roadmap we cover the most promising material systems in recent applications of quantum technologies. These not only include "obvious" material classes like silicon, diamond, rare earth-doped crystals and 2D-materials, but also fields where materials play a supportive or enabling role, as for trapped ions or neutral atom arrays.

Trapped ions form one of the most promising avenues for advancing quantum computing and other technologies. The universal control of trapped ion qubits remains an exciting area of growth and requires precision control of trapping structures and laser/microwave/matter interactions. In a typical device, qubits based on trapped ions are mechanically well-isolated. Yet their control is intimately related to the structural and functional landscape of its nearby materials. Gate architectures and their materials form a key element of these technologies and provide areas of innovation. Quantum enhanced sensing is another attractive area of application using these architectures, giving rise to the possibility of measuring miniscule electrical and magnetic fields in materials and devices. Taking advantage of cryogenic trapping of ions in the mature charge coupled device (CCD) architecture, it is now possible to develop compact high fidelity quantum CCDs simulators.[1] The range of possible applications will greatly benefit from identifying key materials and engineering challenges in these devices and technologies.

[1] Fachrichtung Physik, Universität des Saarlandes, 66123 Saarbrücken, Germany, Email: christoph.becher@physik.uni-saarland.de
[2] Division of Physics and Applied Physics, Nanyang Technological University, Singapore 637371. Email: wbgao@ntu.edu.sg
[3] Department of Physics, Northeastern University, Boston, MA 02115. Email: s.kar@northeastern.edu

Rare earth ions in suitable crystalline matrices can also be exploited for various quantum technologies. Their unusual property that optically active 4f transition levels in these systems are naturally protected by the filled outer shells enable them to be naturally isolated from surrounding perturbations. These systems often possess ultranarrow optical transition linewidths and long coherence times at low temperatures. Transitions in Er based systems arise at or near the 1500 nanometer telecom wavelength, making them attractive for coupling with existing technologies. However, their weak emission rate means additional signal enhancement architectures are needed for single ion readouts.[2] This system holds the promise to address niche problems in computation, memory, and metrology, challenges in controlled manufacturing of rare earth qubits beyond the laboratories will need to be addressed.

With the development of optical tweezing technology, the possibility of harnessing ultracold arrays of neutral atoms have also become a similarly exciting venue for developing quantum logic devices.[3] Advances of this field has largely leveraged the exquisite control of individual neutral atoms or molecules made possible by tightly focused lasers. With the tremendous progress in micro/nanofabrication techniques over the past decades, this process can be replicated to enable trapping of arrays of individual atoms. These atom arrays form a powerful method for simulating emergent phenomena and hold the potential for a variety of quantum technologies.

By virtue of its very advanced engineering history, doped silicon will always remain an attractive material for quantum technologies, from a commercial scaling and high-quality material availability point of view. Isolated phosphorus donors in isotopically pure $^{28}$Si form attractive qubits with spin coherence lasting hundreds of microseconds. Both electron and nuclear spins associated with the dopant P could be candidates for quantum computing applications. Low frequency noise in these systems appear to be extremely low, making them highly attractive for electronically accessible quantum information science. Recent advances in STM-assisted hydrogen de-passivation lithography[4] has enabled engineering of P atoms in the Si matrix with atomic precision – making scalable engineering of Si:P qubits a viable scalable qubit in the near future. Developing these and other high-precision engineering techniques would be a matter of great interest to the quantum technology community.

Single defect centers in wide-band gap semiconductors, such as diamond or SiC to name the most prominent ones, have received particular attention for quantum technology applications due to their often unique combination of long spin coherence times, homogeneous and narrow optical transitions and bright single photon emission. The solid-state nature of their hosts enables their integration into micro- and nanometer-sized structures for enhanced coupling to optical fields or sensing applications. The benefit of established solid-state technologies lies in the option of miniaturization and scalability, and hence attractiveness for applications in fields where size, weight and power consumption are relevant. On the other hand, undesired, accompanying material defects, local strain fields, surface damages or imperfect reconstructions, nuclear spin baths and many more pose a limitation to the advantageous properties of the defect centers and call for improvement in material fabrication and processing. Within this field, the benchmark is set by the negatively charged nitrogen-vacancy (NV$^-$) center in diamond: NV-based quantum sensors defined new application frontiers including biology and condensed matter physics and pioneering proof-of-principle demonstration of quantum network and memory-based communication nodes have been reported. Along with these advancements in diamond color centers, both the optical and spin properties of silicon carbide emitters are optimized by exploiting the matured nanofabrication techniques, optimized quantum control protocol, and material engineering.[5]

In recent times, the spectacular progress in the high-quality synthesis and manipulation of 2D materials, and scalable engineering of defects, impurities and excitonic states are opening up new pathways for niche applications in quantum technologies.[6] Owing to their 2D nature, qubits hosted are exposed to dramatically reduced number of spurious decoherence sources in their immediate vicinity, and hence enhances the possibility of high-coherence-time qubits. Defects in high bandgap 2D materials such as h-BN are have received significant attention as single-photon emitters, and there is a concerted effort by many groups to predict other defects and 2D materials which might possess the right combination of transition energies and coherence timescales of other quantum technologies. This material system can be considered a fresh entrant in the quantum information science field, and hence many theoretical predictions exist. It remains to be seen if, beyond methods for developing high-quality 2D materials, engineering specific types of defects can also be sufficiently advanced, so as to meet the needs of various quantum technologies.

Photons are an integral part of a variety of quantum technologies, and single photon detection is crucial for efficient quantum information science.  All protocols for quantum communication rely on the faithful detection of single photons; figures of merit are the detection efficiency determining e.g. the rate of secret key generation and the signal-to-background/noise ratio determining the communication range and fidelity of quantum state transmission. Only the advent of ultra-low dark count and highly efficient superconducting detectors[7] made possible demonstrations of quantum key distribution (QKD) over hundreds of kilometers. On a more fundamental side, both properties enable closing loopholes in fundamental tests of quantum mechanics (via Bell-type inequalities) which in turn pave the way for advanced quantum communication protocols like (measurement-) device-independent QKD. Furthermore, superconducting single photon detectors (SSPD) are an enabling technology for chip-based photonic quantum computing: here the quantum state of photons has to be determined fast and with high fidelity to allow for feed-forward operations in measurement-based "one-way" quantum computing. The challenges on the material side are the optimization of superconducting properties (still less noise, higher efficiency over broad wavelength ranges, less dead time etc) but also the hybrid integration of SSPDs with materials for photonic integrated circuits (silicon, silica, lithium niobate etc).

Recently, Giustino *et al.*,[8] have published a Roadmap on quantum materials. In the quantum world, an overlap of quantum technologies with quantum materials inevitable. Hence, topics that were sufficiently covered there (*e.g.* superconducting and topological materials) are intentionally not addressed in detail in our present Roadmap. In general, we have attempted to keep our discussions distinct from this and other recent review on quantum materials, by keeping our focus on issues that pertain specifically to identified quantum technologies.

The field of quantum technologies has witnessed dramatic evolutions over the last two decades. Much of the progress was driven by ingenious ideas and concepts and experimental demonstrations ranging from proof-of-principle to first working prototypes. Further development in the field, independent of the nature of the physical quantum bit systems, will hinge on material development for improving optical and spin coherence, reducing inhomogeneous distributions of physical properties, interfacing quantum bits to the environment and enabling scalability. Progress is required in a large range from material composition and purity, reduction of residual defects, targeted doping and isotopic purification, enhanced surface definition and optimized interfaces at material boundaries to advanced material processing technologies like nanoscale etching, hybrid material integration, robust interfacing to optical and electrical industry-norm connectors, integration with control electronics and packaging into rugged devices. As all individual steps and methods become known or ready to be developed, the path to quantum technologies in field- and everyday-applications will be widely opened.

## 2. Trapped-ion Quantum Technologies

Christian Marciniak, Institut für Experimentalphysik, 6020 Innsbruck, Austria
Thomas Monz, Institut für Experimentalphysik, 6020 Innsbruck, Austria & AQT, 6020 Innsbruck, Austria

Status

Trapped charged atoms are an established platform spanning the breadth of quantum technologies from metrology and quantum communication, to simulation and quantum computing. The strong case for this platform is the near-perfect environmental decoupling of identical constituents. Trapped ions have historically led quantum technologies in terms of performance fidelities, with unmatched state preparation and readout error rates. They hold the center position among leading technology platforms in terms of duty and measurement cycle, as well as - for quantum computing - qubit number. The following roadmap for the ion-trap platform will focus on quantum computing, as many aspects are common between quantum technology pillars.

Qubits employed in ion-trap quantum computing can be grouped based on encoding. Information is stored either in metastable excited or in ground states of the respective ion. Qubits employing metastable states have been historically amongst the first realisations of ion-trap quantum information registers. The metastable state typically has a lifetime of about 1s. Gate operations with error rates much smaller than 1% have been achieved using such laser-manipulated qubits with gate times of about 100 µs. The lifetime of the computational states in ground state qubits on the other hand is practically infinite, and transitions are mediated via optical Raman or microwave-driven dipole interactions. In particular microwave-based approaches have recently demonstrated gate error rates as low as ~$10^{-4}$ for single qubits, and ~$10^{-3}$ for two-qubit entangling gate operations.

In quantum information processing these qubits are most-often trapped in linear configurations using various electrode designs. Traps can be grouped into 3D traps (usually based on few elements), and surface traps that are commonly manufactured adopting technologies from the semiconductor industry to produce hundreds or more individually-controlled electrodes. 3D trap designs are technologically mature, and can stably store linear arrays of 50 to 100 ions, while high-fidelity, individual control is limited to less than 50. Surface (or charged coupled device - CCD) traps on the other hand are a new technology that has emerged from technical considerations when scaling the platform to significantly larger qubit numbers. 2D ion structures in 3D traps are pursued primarily for analog quantum simulation, where hundreds of ions with global control are well-established.

Trapped-ion systems are naturally suited as network matter-nodes, and are being actively integrated with photonic links for quantum communication and distributed computing architectures. Modularization, standardization, and pioneering integration of quantum computing hardware into high-performance computing and data centers is still in its infancy.

Current and Future Challenges

The primary challenges for trapped-ion technologies going forward are overcoming the limitations of current gate schemes in intrinsic infidelity, required time, and maintaining the control demonstrated as system size is scaled up. Conceptual problems that have been solved for smaller registers are then becoming major engineering challenges whose solution is likely to require novel approaches.

It is widely believed that linear chains in a single potential cannot be extended significantly past hundreds of qubits. Segmented and CCD traps allow for much higher qubit density, but suffer from so-called anomalous heating which often presents the dominant error source. All architectures face the constraints that collisions with background gas particles impose. These collisions become increasingly likely as the qubit count grows, reducing both the duty cycle, and the longest time a single computation can last.

Today, one of the major roadblocks is the question of how to scalably provide control as register sizes grow. Light-based, single-site addressing exists, but no approach has emerged as the superior strategy for large-scale deployment. Addressing crosstalk already presents a bottleneck in contemporary quantum computers. Alternatively, segmented or surface traps are frequently manufactured with dedicated interaction zones with either integrated, microfabricated optical or electronic addressing sites, or stationary processing zones through which ions are transported electrostatically. However, these transport-based approaches typically perform slower than macroscopic addressing approaches. For some ion species, integration of optical components is further complicated by solarization caused through short-wavelength radiation. Maintaining low temperatures in the relevant modes is a concern that is becoming increasingly pressing. While laser cooling gives access to near-perfect ground state occupation this typically relies on techniques that scale poorly in necessary time or implementation effort as the register grows. It is ultimately unlikely that a single processor will provide sufficient qubits for large-scale computation. Enabling distributed computing via for example photonic links is currently hindered by low-efficiency or poorly-scalable light collection, non-deterministic photon generation, and propagation losses which lead to repetition rates far below individual processor capabilities.

Metastable qubit implementations are restricted to comparatively shallow circuits with current gate schemes despite approaching the natural limit in coherence time. While Raman or microwave qubits can overcome the lifetime restriction, they face mechanical and electrical engineering challenges in maintaining interferometric stability. It is clear that the number of gates within the coherence time needs to significantly increase in the coming years, which is likely to require novel approaches to performing gates.

### Advances in Science and Technology to Meet Challenges

The increasing demands on vacuum quality with growing qubit number are likely only to be met by moving from room-temperature setups to cryogenic environments. Management of vibrations through cryo equipment will be particularly important for high-fidelity Raman operation. Integrated waveguide and conductor solutions can help to address these new challenges. Scalable distribution of electric and optical waveguides needs to respect the available space, the low-field-noise environment required, and thus likely require advanced switching and multiplexing capabilities. Maintaining low latencies and minimizing heat conduction is likely going to require cryo-compatible, in-vacuum control electronics. As the heat load through active or dissipating elements and higher line density to the outside of the cryostat increases, so does the need for higher cooling power. At the same time, these developments need to be cognisant of the availability of He-3/He-4, whose global and local supply in the face of rising demand may prove a critical limitation rather than just a financial burden.

Cryogenic temperatures are also required to combat anomalous heating in surface trap architectures. While poorly understood, evidence is compelling that this heating originates at the electrode surfaces. Material science and surface treatment technologies need to advance to significantly reduce its magnitude. At the same time, these materials need to be amenable to integration with non-telecom wavelengths, and allow for high slew rate and large arcing voltages to enable faster transport operations of ions.

Increasing gate speed across all architectures towards nanosecond scales will mean overcoming the limitations of Coulomb-mediated interactions. Ultrafast Rydberg- or photon-kick-based gates have been realized, but fault-tolerant operation is yet to be demonstrated. Faster gate speeds put even higher demands on shuttling times, as well as photonic links which are unlikely to be met without integrated electronics, and off-the-shelf high-fidelity single-photon conversion, collection, and detection.

Seeing that the majority of qubits in ion trap-based quantum computing are currently manipulated by lasers, more reliable, industry-grade lasers will face a growing demand, including the availability of equipment to stabilize them without intervention for months rather than days. Scalable generation, routing, and distribution of atomic-physics-quality electromagnetic radiation will require advances in massively-integrated waveguide technology, amplification, and low-loss high-extinction switches.

### Concluding Remarks

Trapped-ion quantum technologies face many important challenges on their path to large-scale deployment. The prospect of large quantum machines running on naturally identical qubits which offer themselves readily for photonic interconnects in distributed computing or communication is a strong drive in this pursuit. Fortunately, the trapped-ion platform shares many challenges on this trajectory with other leading platforms, from cryostat development to demands on electronics, or large-scale integration. Not only continuing to share, but increasing synergies between platforms will benefit the quantum technology community as a whole.

### Acknowledgements

*We gratefully acknowledge funding from the EU H2020-FETFLAG-2018-03 under Grant Agreement no. 820495. We also acknowledge support by the Austrian Science Fund (FWF), through the SFB BeyondC (FWF Project No. F7109), by the Austrian Research Promotion Agency (FFG) contract 872766 and 884471, and the IQI GmbH. This project has received funding from the European Union's Horizon 2020 research and innovation programme under the Marie Skłodowska-Curie grant agreement No 840450. T.M. acknowledges funding by the Office of the Director of National Intelligence (ODNI), Intelligence Advanced Research Projects Activity (IARPA), via US ARO grant no. W911NF-16-1-0070 and W911NF-20-1-0007, and the US Air Force Office of Scientific Research (AFOSR) via IOE Grant No. FA9550-19-1-7044 LASCEM.*
*All statements of fact, opinions or conclusions contained herein are those of the authors and should not be construed as representing the official views or policies of the funding agencies.*

# 3. Rare Earth Ions for Optical Quantum Technologies

John G. Bartholomew[1] and Philippe Goldner[2],
[1]Centre for Engineered Quantum Systems, School of Physics & Sydney Nanoscience Institute, The University of Sydney, Sydney, Australia, [2]Chimie ParisTech, CNRS, PSL University, Institut de Recherche de Chimie Paris, Paris, France

### Status

Crystals containing rare-earth ions have a rich history in the field of spectroscopy dating back to the mid-19th century. As purification techniques, laser technology and material science improved there has been a natural progression from studying the fascinating mechanics of these systems to developing optical technologies. In particular, the pursuit of classical information technologies generated physical models and spectroscopic techniques that established a solid foundation for applying these materials to solve challenges in quantum science. The key appeal of rare-earth, from single ions right through to large ensembles in stoichiometric crystals, is their multiple coherent degrees of freedom. The most prominent is extremely narrow 4f-4f optical transition homogeneous linewidths at liquid helium temperatures; as narrow as 73 Hz (or coherence lifetime $T_2$ = 4.4 ms) in $Er^{3+}$:$Y_2SiO_5$ single crystals [1]. Such low broadening is possible because the 4f electrons are shielded from lattice perturbations by closed outer electron shells. In addition, rare-earth ions can also possess highly coherent electron or nuclear spin degrees of freedom that can be optically controlled. At suitable temperatures and magnetic fields, these spins can show coherence lifetimes $T_2$ well into the 10s of ms range, and up to 6 hours for the nuclear spins of $Eu^{3+}$:$Y_2SiO_5$ [2,3].

These distinct properties led to the early recognition that rare-earth crystals were extremely well suited to quantum memory and processor technologies [2]. Rare-earth ion technology is a leading quantum memory platform with demonstrations of long storage times using spin degrees of freedom, high retrieval efficiency, multimode storage, quantum memories entanglement and light matter teleportation [4]. Developing in parallel were initial proposals and experiments related to quantum processing based on ensembles of ions [2]. Recently, advances in single ion detection, especially by coupling to micro- or nano-optical resonators, have led to single spin coherent control and single shot readout, opening the way to optically addressable spin qubits [5,6]. The combination of quantum memories and qubit processing nodes could allow quantum information to be distributed over large distances through optical quantum repeater links. Rare-earth ion technology provides the basis for generating, storing, processing and converting information at the quantum level (see Figure 1), which are critical elements required for an integrated quantum network.

An additional strength of the rare-earth ion platform is its versatility. For example, the 4f-4f optical transitions in different rare-earth ions cover a broad spectral range, from the infrared to the UV, including the important telecom wavelength at 1.55 μm found in erbium. These ions can be introduced in various insulating crystals, glasses, and semi-conductors either by doping during synthesis or implantation. This provides access to a very large number of host-dopant combination, which can be leveraged not only as bulk materials but also as micro- or nano-structured materials such as nanoparticles or thin films [7]. Although many of the achievements place the rare-earth-ion system among the best systems for quantum information applications, these successes are still limited by material challenges. Therefore, the removal of these restricting factors is truly an exciting prospect.

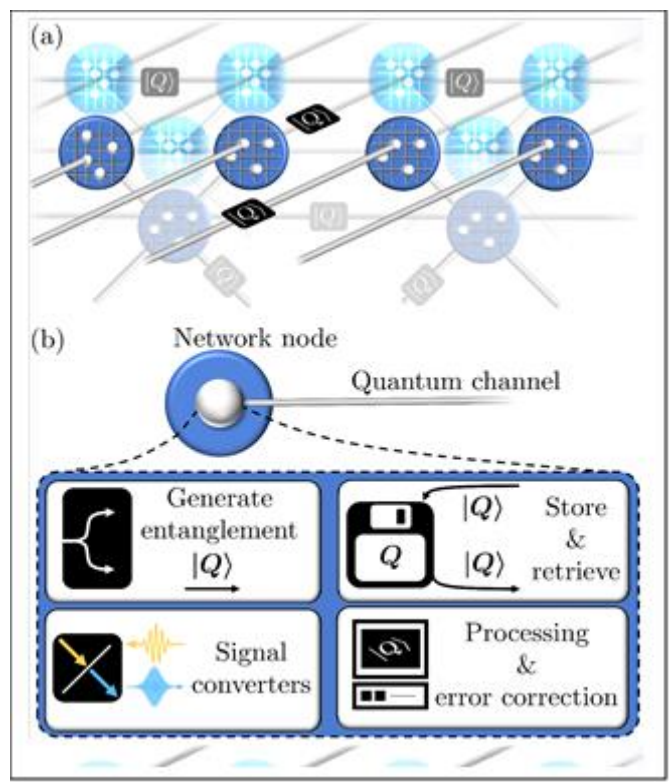

Figure 1. Enabling technologies for quantum networks. Quantum information networks will continue to grow in complexity (a) to ultimately allow quantum states to be distributed among numerous small scale local network (blue spheres) and over longer distances. To realize robust and scalable quantum networks, each network node (white sphere) will need to incorporate a suite of technologies (b). Such technologies include sources of entanglement, quantum memories for network synchronization and repeater stations, converters and transducers to act as adapters for quantum technologies operating in different physical regimes (e.g. microwave and optical), and quantum computers to process information and perform error correction operations. Rare-earth ion crystals are among the leading material systems to realize the varied devices that are critical to quantum network performance. Translating current work to deployable technologies in large, complex networks will be accelerated through a deeper understanding of material limitations and new breakthroughs in material synthesis and fabrication.

Current and Future Challenges

A first important research topic is to identify and grow improved bulk crystals. Many currently used crystals limit quantum device performance through the presence of abundant nuclear spins. For example, in $Y_2SiO_5$, $^{89}Y$ has a 1/2 nuclear spin with 100% abundance and the magnetic noise caused by this spin bath can ultimately limit coherence lifetimes, especially for electron spins. This problem is even worse in hosts like $YVO_4$ or $LiNbO_3$, which contain nuclear spins with large magnetic moments. In addition, nominally undoped yttrium-based crystals contain unavoidable rare-earth ion impurities at a 0.01 – 1 parts per million level, which is detrimental to single ion qubit devices. Several crystals with heavily reduced nuclear spin concentration and ultra-low rare-earth background impurities show promising properties such as $Er^{3+}$-doped $CaWO_4$, Si, or $TiO_2$ [8]. A challenge, however, is that these materials do not have substitutional sites with well-matched size and charge for trivalent rare earth ions, which may lead to additional defects and strain and in turn potential sources of decoherence.

Integrating rare-earth ions with photonic or microwave resonators presents another important materials challenge (Figure 2). Such coupling is essential in ensemble and single ion devices for high-efficiency and high-fidelity operations. For example, ion-resonator coupling is critical to overcome the weak rare-earth ion optical oscillator strength (» $10^{-6}$) to achieve high efficiency single ion qubit detection and control. An ideal host would combine low optical or microwave waveguide and cavity loss with optimised rare-earth ion properties like narrow homogeneous and inhomogeneous linewidths. This can be at least partly obtained in crystals well developed for integrated optics like $LiNbO_3$ or silicon where RE ions can be introduced by doping of implantation. Structure can also be fabricated directly into crystals well suited to RE ions like $Y_2SiO_5$ [5]. Hybrid designs with surface resonators or open cavities are also available and could provide optimised optical and RE properties at the same time [6,9].

Attempting to increase the coherent ion-resonator coupling rate in integrated systems by minimising the resonator mode volume increases the ions' proximity to interfaces and the likelihood of increased decoherence. This key trade-off is common to all architectures for miniaturised devices including resonators that are milled, etched or deposited onto the surface of bulk crystals. Efforts to overcome this trade-off include mitigating near-surface electric field perturbations by using rare-earth ions in non-polar site symmetries in which permanent electric dipole moments vanish for all electronic levels. This reduces electric field broadening at the expense of one avenue for two-qubit gate operations and the ability to frequency tune using electric fields.

Another option is to improve material synthesis, fabrication, and post-processing, concentrating specifically on the crystal interfaces. The synthesis of rare-earth ion nanoparticles or thin films offer a way to study the optimisation of these material improvements as well as appealing alternate architectures for devices, such as coupling to open micro-cavities. However, RE ions in nano-materials generally show degraded properties, most notably optical $T_2$, compared to single crystals. Efforts to circumvent these problems are currently focusing on improving synthesis and post-processing of nanomaterials [7].

High-resolution and controlled spatial localization could be achieved with implantation in hosts with low background RE impurities. As RE ions are heavy elements it is expected that significant damages occur during implantation. It has been shown that annealing can at least partially cure these defects, leading to relatively narrow lines [10]. It is however unclear to what extent implanted ions properties can approach doped ions ones.

### Advances in Science and Technology to Meet Challenges

A fundamental requirement to efficiently optimize materials is to be able to link structural information with relevant properties. In the case of rare-earth ion coherence lifetimes, accurate models have been developed for bulk crystals in regimes where the rare-earth or host electron spins are the main source of dephasing. The role of defects, impurities and residual disorder is less clear and often difficult to relate to structural analyses but is still likely to be a limit at low doping concentration or in spin-free hosts. For ions close to surface, charge noise is a likely cause of excess dephasing and spectral diffusion compared to the bulk ions but direct experimental evidence is yet to be demonstrated. In the same way, the relatively short optical $T_2$ in nanoparticles and thin films are likely to be due to charge noise at the surface and also within the lattice as a result of a higher concentration of defects such as oxygen vacancies. A deep understanding of these different results is likely to emerge from comprehensive studies that involve a range of materials and techniques like XRD, electron microscopy, chemical analysis, electron paramagnetic resonance, luminescence of defects, and their correlation with RE spectroscopy. From this understanding, optimized synthesis of materials and post-processing can be designed starting from the broad range of techniques available for bulk and nanoscale materials.

In the case of implanted samples or nanomaterials, low rare-earth concentration or material volume results in low signals, which makes rapid feedback towards material development more difficult. Here, experimental throughput could benefit from the progresses in low temperature microscopy and high sensitivity superconducting nanowire detectors. An even more difficult investigation is single ion spectroscopy, which currently requires material specific nano-fabrication techniques. The development of easy to operate, high throughput experimental techniques such as open optical micro-cavities or samples integrated with superconducting detectors could solve this issue for nanomaterials and bulk sample thinned down to micron-thin membranes.

Hybrid approaches for rare-earth integration of into photonic structures could be improved in several ways. As stated above, materials coherence properties at the nanoscale should be first improved. Second, high quality light waveguides and resonators adapted to rare-earth wavelengths in the near IR or visible regions would enable exploiting new rare-earth transitions beyond the silicon compatible $Er^{3+}$ telecom wavelength. Another technology of potentially broad applicability is laser written photonic structures, which seem to affect rare-earth coherence only to a small extent [11].

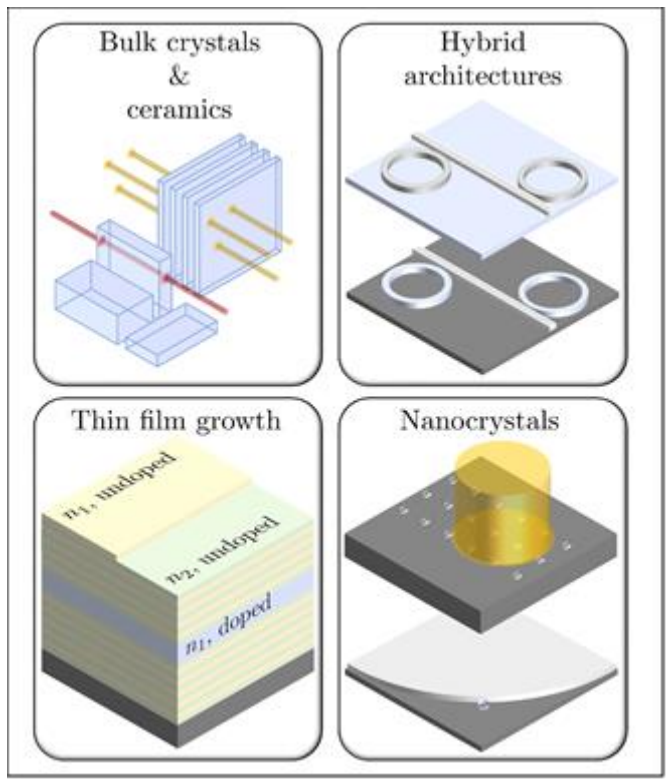

Figure 2. Multiple avenues for technology development. The rare-earth ion crystal platform has the versatility of enabling devices for quantum information science in both the ensemble and single ion regimes. Continued progress is required across a number of different architectures to fully leverage their multiple high coherent degrees of freedom. Bulk crystals and ceramics offer opportunities for large capacity long-term quantum storage. Hybrid architectures combine high performance photonic layers with bulk rare-earth ion materials and devices. Thin film growth dramatically increases the large-scale fabrication possibilities in this material system and may allow new architectures yet to be realised through traditional bulk crystal growth. Nanocrystals represent the ultimate miniaturization of rare-earth ion material systems and offer alternate device implementations as well as new insights into the decoherence mechanisms that ultimately govern device performance.

## Concluding Remarks

Rare earth ions in crystalline hosts are promising systems for optical quantum technologies, showing optical and spin transitions with some of the longest coherence lifetimes demonstrated in the solid-state. Further exploitation of these exceptional properties towards integrated quantum photonic architectures leveraging the ensemble and single ion regimes requires significant improvement in material engineering, as coherence lifetimes can be reduced by order of magnitudes compared to bulk materials at the nanoscale. Such developments should be guided by a comprehensive understanding of the dephasing mechanisms involved. Otherwise, novel high quality photonic structures that can couple with RE crystals are needed. As the latter can be obtained in different forms (bulk, nanoparticles, thin films) or and operate at different wavelengths, a variety of successful combinations can be expected for different applications. In this respect, the search for new material platforms, such as rare-earth molecular crystals, or through atom-scale precision site engineering, could enormously broaden the scope of possible systems.

**Acknowledgements**

The authors would like to acknowledge the support of the Australian Research Council Centre of Excellence for Engineered Quantum Systems (EQUS, CE170100009), and of the European Union's Horizon 2020 research and innovation programme under grant agreement No 820391 (Square).

## 4. Atom Arrays for Quantum Simulation

Huanqian Loh
National University of Singapore, Singapore

**Status**

Ultracold atoms individually trapped in arrays of optical tweezers have rapidly emerged as a powerful platform for quantum simulation of many-body physics. These atom arrays offer scalable single-qubit control over both the internal spin state and the external motional quantum state. They can also be generated in arbitrary geometries in one, two, or three dimensions with programmable doping and tunable interactions. These combined features of scalability and programmability render them highly attractive as synthetic materials for analog quantum simulation, where the atom arrays are set up to obey the same Hamiltonians used to model materials.

To generate the atom arrays, atoms are first stochastically loaded from optical molasses into an array of optical traps. These optical traps are formed by sending the trapping laser through either a spatial-light modulator or acousto-optic deflectors, and then focused onto the atoms using a high numerical-aperture microscope objective. At this stage, the filling fraction is usually limited by collisional blockade to be about 50-60%. The filled trap positions are imaged using atom fluorescence and then fed back in real time to acousto-optic deflectors that generate one or more moving tweezers to rearrange the atoms by dragging them to target sites (Fig. 1). Such dynamic reconfiguration has enabled the creation of defect-free atom arrays in user-defined geometries ranging from honeycomb lattices [1] to three-dimensional fullerene-like arrays [2].

To map the defect-free array onto a given model Hamiltonian, tunable interactions need to be induced between the singly-trapped atoms. Exciting atoms to Rydberg states (principal quantum number ~ 60-80) emerge as a natural route for inducing tunable interactions in atom arrays, as the Rydberg interactions $V/h$ can remain on the scale of 1-10 MHz for typical tweezer spacings of several microns. Such strong Rydberg interactions can be induced via one of the following ways:
(1) coherently driving the array of atoms between the ground state $|g\rangle$ and a Rydberg state $|r\rangle$ with a laser resonant to the ground-Rydberg transition frequency for a single atom;
(2) transferring the atoms to the Rydberg state using stimulated Raman adiabatic passage (STIRAP) and subsequently driving the atoms between two Rydberg states $|r_1\rangle$, $|r_2\rangle$.

The first method induces van der Waals interactions between atoms in the same Rydberg state. These interactions scale as $C_6/r^6$, where $r$ is the distance between atoms. When an atom array is uniformly illuminated by a resonant Rydberg laser of Rabi frequency $\Omega$, Rydberg blockade occurs, where atoms within a given blockade radius $R_b=(C_6/(\hbar\Omega))^{1/6}$ of a Rydberg atom are prevented from being excited to the Rydberg state as their transition frequencies are shifted off resonance. Mapping the ground state and Rydberg state onto $|\downarrow\rangle$ and $|\uparrow\rangle$ respectively allows one to implement an Ising-type Hamiltonian with the atom array [3]. By ramping the detuning adiabatically across the Rydberg transition resonance, crystalline phases of $Z_n$ symmetry have been observed, where $n$ refers to the number of neighboring blockaded atoms and is controlled by varying the array spacing relative to the blockade radius [4]. Beyond phase transitions, the atom array is also well suited for studies of spin dynamics, as the quantum states of individual atoms in the array can be detected directly through atom fluorescence. For instance, dynamical signatures of quantum many-body scars have been observed by bringing the atom array to a rapid quench and seeing the atoms exhibit long-lived coherent oscillations between the states $|\downarrow\uparrow\downarrow\uparrow...\rangle$ and $|\uparrow\downarrow\uparrow\downarrow...\rangle$ [4].

The second method of Rydberg excitation induces dipolar interactions between atoms in the two different Rydberg states. These dipolar interactions are both long-range (scaling as $C_3/r^3$) and anisotropic. Coupling between the two Rydberg states is typically achieved using a microwave field. Spin models like the XY and XXZ spin Hamiltonians can be engineered by mapping the two Rydberg states onto $|\downarrow\rangle$ and $|\uparrow\rangle$ [3]. For instance, the Su-Schrieffer-Heeger model, initially proposed to explain the high conductivity of doped organic polymers, was realized for a synthetic one-dimensional atom chain by arranging the Rydberg atoms such that next-nearest-neighbor hoppings are suppressed by the anisotropy of the dipolar interactions, whereas nearest-neighbor couplings alternate between a strong and a weak value (Fig. 2) [5]. This led to an observation of a symmetry-protected topological phase for interacting bosons [5].

So far, the above methods of inducing tunable interactions involve direct excitation of atoms to the Rydberg state. While these Rydberg states boast large interaction strengths, their lifetimes are typically limited to hundreds of microseconds. A complementary method of Rydberg excitation involves admixing the ground state with a small component of the Rydberg state by using an off-resonant dressing laser, where the Rabi frequency is small compared to its detuning. This Rydberg dressing allows for interactions between two long-lived spin states in the electronic ground state and offer possibilities for realizing more sophisticated spin Hamiltonians [3].

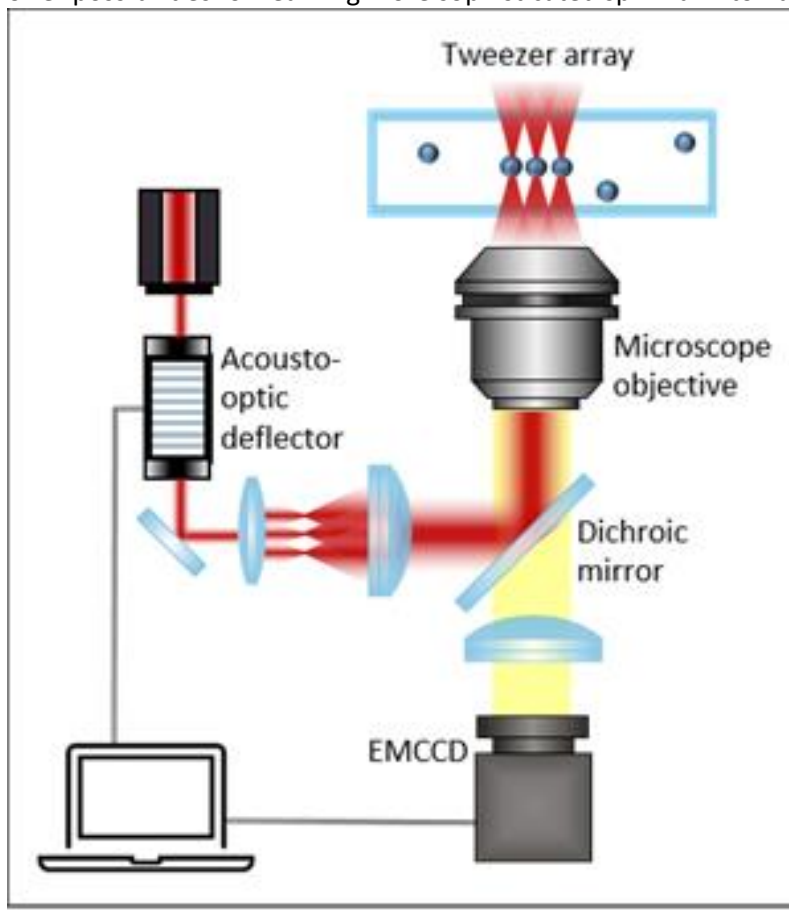

Figure 1. An acousto-optic deflector (AOD) generates multiple beam deflections, which are imprinted as an array of tweezers onto the atom plane via a microscope objective that simultaneously collects atom fluorescence on an electron-multiplying CCD (EMCCD) camera. Real-time feedback from the EMCCD to the AOD generates a defect-free atom array.

**Current and Future Challenges**

Presently, state-of-the-art quantum simulation experiments with atom arrays involve hundreds of singly-trapped atoms [1]. Scaling the size of defect-free arrays up to thousands of atoms remains a significant challenge. The array size is typically limited by tweezer laser power, which is in turn related to a combination of factors like the minimum trap depth required to load and image atoms, atom polarizability at the chosen tweezer wavelength, and availability of high power lasers.

Concurrently, as the array size increases, high-fidelity rearrangement of atoms to generate defect-free arrays becomes more difficult. This is because larger arrays tend to require longer reconfiguration times, yet atoms have a finite trap lifetime due to collisions with the background gas, which causes an exponential decay in the probability of a defect-free target array at the end of rearrangement. Therefore, efficient algorithms that minimize the rearrangement time or techniques that suppress the atom loss mechanisms need to be developed.

Besides scalability, establishing a higher degree of control over the atom-array platform would help push the frontiers of quantum simulation experiments. These include better programmability of interactions, higher-fidelity quantum state preparation and readout, and improved spin coherence. Some of these challenges can be tackled by using different atomic species, while others can be mitigated through technical improvements. For instance, where atom-array experiments involve two-color Rydberg excitations, the ground-Rydberg coherence times are limited by Rydberg excitation laser phase noise, spontaneous emission from the intermediate state, and Doppler effects from finite atom motion. These can be improved respectively by cavity filtering of laser phase noise, using high-power Rydberg lasers to compensate for larger detunings from the intermediate state, and cooling atoms to the motional ground state.

**Advances in Science and Technology to Meet Challenges**
To scale up atom arrays, efforts have been made to build high-power lasers at conventional tweezer wavelengths by frequency doubling fiber-amplified lasers. A complementary approach is to reduce the trap depth required to load atoms. For alkali atoms, D1 light has been shown to not only cool atoms efficiently when loading them into shallow traps, but also yield enhanced loading probabilities as high as 90% by driving atoms to a repulsive molecular state [6]. Combining these techniques with D1 magic wavelength tweezers further increases the overall scalability by reducing the trap depth used when imaging atoms [7].

For high-fidelity generation of large defect-free arrays, the atom array can be placed in an ultra-high-vacuum-compatible cryostat to improve the atom lifetime [8]. Such a cryogenic setup can be combined with parallel rearrangement algorithms and the aforementioned methods for enhanced loading probabilities to reduce the rearrangement time. The cryogenic temperatures would also suppress black-body radiation-induced transitions, which can lead to longer Rydberg lifetimes and Rydberg-dressing coherence times.

Beyond the alkali atoms typically used in atom-array experiments, alkaline earth atoms are recently pursued for their increased coherence and high-fidelity state preparation and readout methods [9]. Multiple atomic species can also be combined in a single setup to explore alternative ways of inducing programmable interactions, such as by combining the pre-cooled atoms to form ultracold dipolar molecules in tweezer arrays [10]. Such singly-trapped molecules, whose electric dipole moments can be manipulated in the ground state using both electric fields and microwaves while avoiding chemical reactions with other molecules, offer rich possibilities for realizing exotic Hamiltonians and probing spin dynamics with long coherence times.

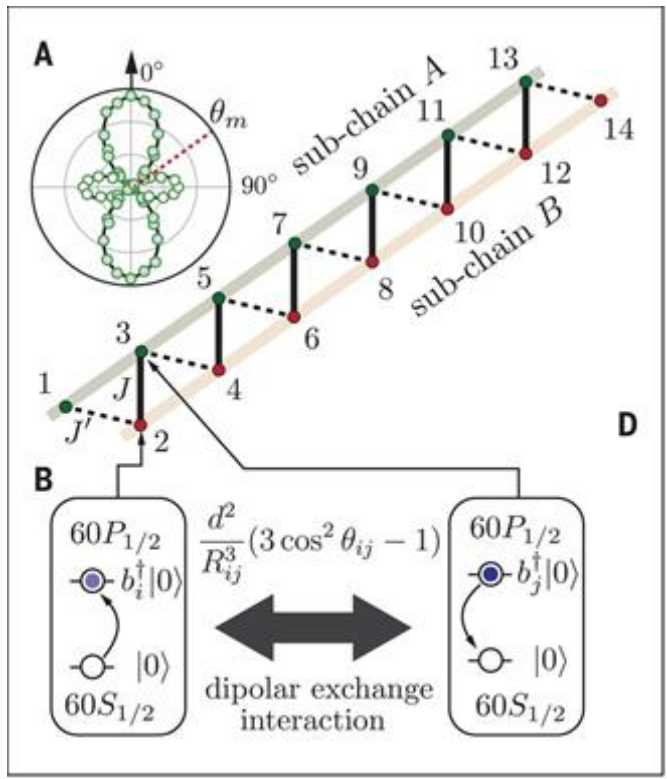

Figure 2. Quantum simulation of the Su-Schrieffer-Heeger model, realized by (a) arranging atoms along two sub-chains at an angle $\theta_m$ to cancel out next-nearest-neighbor interactions and (b) mapping two distinct Rydberg states, $60S_{1/2}$ and $60P_{1/2}$, onto two spin states. From [5]. Reprinted with permission from AAAS.

**Concluding Remarks**

Atom arrays as synthetic materials are one of the leading platforms for analog quantum simulation of many-body and dynamical physics. Further improvements in scalability, control, and programmability will not only push the frontiers of quantum simulation, and also make atom arrays highly competitive for quantum information processing in general.

**Acknowledgements**

We acknowledge support from the National Research Foundation (NRF) Singapore (grant no. NRF-NRFF2018-02) and the Ministry of Education, Singapore.

# 5. Donor-Based Silicon Quantum Technologies

Elizabeth Marcellina[1], Kuan Eng Johnson Goh[2,3], Teck Seng Koh[1], Bent Weber[1]

[1]Nanyang Technological University
[2]Agency for Science, Technology, and Research (A*STAR)
[3]National University of Singapore

Status

Silicon has been the cornerstone of the modern semiconductor industry for over six decades, and, since the seminal proposal by Kane in 1998 [1] (Figs. 1A-C), has also become a promising platform for quantum technologies [2,3]. Owing to impressive advances in semiconductor device physics and engineering, the past two decades have seen key demonstrations of quantum processor prototypes, including those in donor-based architectures [3-11] (Figs. 1D-I).

The main motivation for utilizing silicon as a host material for quantum technologies is rooted in the weak interactions of electron (or nuclear) spins with the crystalline host. The resulting exceptionally long spin life- ($T_1$) and coherence ($T_2$) times have earned silicon the nickname of a "semiconductor vacuum" [12], a condensed-matter analogue of cold-atom or trapped ion systems hosted in ultra-high vacuum environments. Indeed, the spin-orbit interaction for electrons is generally considered weak [11], and the already weak contact hyperfine fields in natural silicon ($^{nat}Si$) can be further reduced in isotopic purified $^{28}Si$. This is particularly true for donor spin qubits in silicon, due to their atomic-scale footprint with a tightly confined, spherically symmetric electron wave function.

Electron (ESR) and nuclear (NMR) spin resonance experiments of phosphorus ($^{31}P$) bulk donor ensembles as early as the 1960s and have shown electron and nuclear spin lifetimes ($T_1$) of ~1 and ~10 hours [13], respectively, at low temperature, even in $^{nat}Si$. Recent measurements of donor ensembles in $^{28}Si$ have yielded coherence times $T_2$ = 10 s for the electron spin [14]. Dynamically-decoupled (DD) coherence times of up to $T_{2DD}$ = 3 min have been measured for the nucleus if the donor is in its neutral state [12], or even 180 min if it is ionized [15].

To date, two established technologies exist to realize nanoelectronic devices based on $^{31}P$ donor qubits (Figs. 1F-I). Most of the early landmark demonstrations have been achieved by high-throughput ionimplanted metal-oxide-semiconductor (MOS) compatible device architectures [4-8] (Figs. 1F,H). These demonstrations include electron (ESR) and nuclear (NMR) spin resonance of individual qubits, detected in projective single-shot electron spin read-out. More recently, a CROT exchange gate [4], and even an entangled Greenberger-Horne-Zeilinger three-qubit state of two $^{31}P$ donor nuclei coherently coupled to a single electron spin [7], have been demonstrated.

Alternatively, donors can be placed with an accuracy <1 nm using scanning tunnelling microscopy (STM) lithography [3,9-11] (Figs. 1G,I). In this approach, the qubits themselves are embedded in the silicon single-crystal alongside control electrodes, gates, and single-electron transistor (SET) charge sensors in an all-epitaxial two-dimensional (2D) co-planar device architecture [3,9-11]. The high spatial accuracy in donor placement allows for precision engineering of the strength of the exchange coupling *J* [9], sensitive to both separation and orientation of a donor-qubit pair in the silicon matrix and has subsequently allowed for the demonstrations of a fast (800 ps), high-fidelity (~94%) exchange gate (Figs. 1G,I) [3].

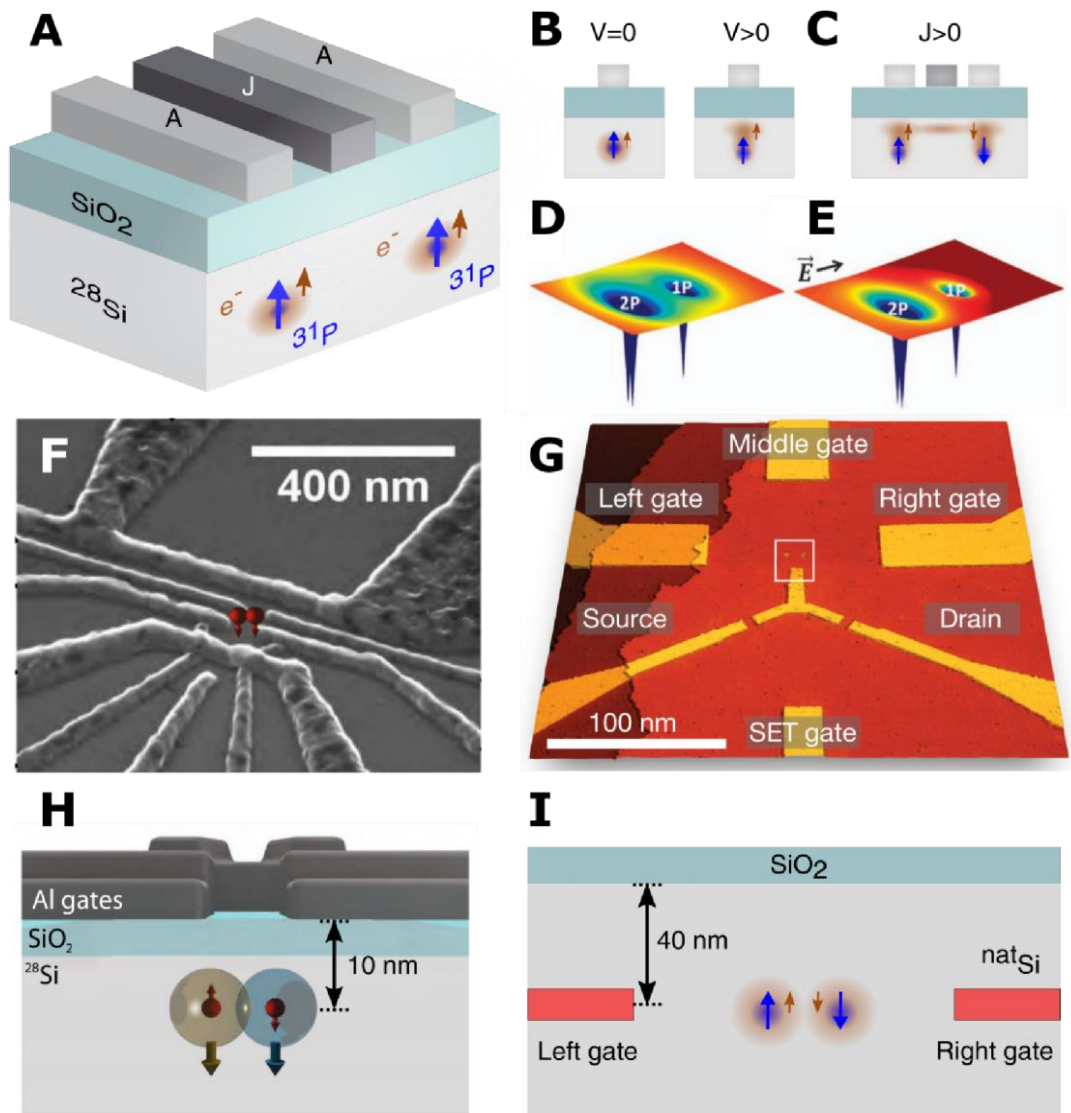

**Figure 1. Si:P Donor-based quantum processors.** (A) Original Kane [1] proposal based on an array of phosphorus donors. Both electron (brown arrows) and nuclear spins (blue arrows) are qubits. exchange coupling, respectively. (B, C) Electron wavefunction underneath *AA-- gate* and s *J*-agates control the hyperfine and nd *J*-gate, respectively. (D) The **exchange** interaction can be further controlled by embedding an asymmetric 1P-2P donor molecule, and (E) tuned by the application of a detuning electric field $\vec{E}$ [*npj Quantum Inf.* **2** 16008]. (F, H) An SEM image and the cross-section of the top-gate stacked of a donor-based two-qubit device, fabricated via single-ion implantation [4]. (G, I) An STM micrograph and a cross-section of a two-qubit device, with phosphorus atoms deposited by STM-lithography [3].

## Current and Future Challenges

The single-most sizable challenge in donor-based quantum computing is to scale to an appreciable number of qubits. Kane's original architecture [1] envisioned linear arrays of donor qubits, precisionplaced at separations od order 10 nm, dictated by the strength of the exchange interaction. Yet, for any technique, it remains challenging to embed $^{31}$P donor atoms into Si accurately, reproducibly, and efficiently towards such scalable arrays. Despite the impressive technological advances in deterministic single-ion implantation [16], ion straggle may inherently limit placement accuracy to the ~10 nm scale. Implanted $P_2$ molecules have allowed for demonstrations of two-qubit gates. However, the separation and orientation of the implanted donor molecule can usually not be controlled, and requires that suitable donor pairs be electrically isolated from large implanted ensembles [4]. At the same time, STM-assisted qubit placement remains limited in throughput and device yield, despite its advantages in accuracy. More so, the stochastic nature of the dissociation and incorporation chemistry of the dopant precursor molecule results in donor number fluctuations at qubit incorporation sites, often resulting in the placement of donor clusters rather than individual donor atoms [3,9]. Alternative precursors for both donors and acceptors have recently been considered, although they have yet to reach comparable maturity with the Si:P based fabrication. To further streamline the STM-based fabrication process for fabrication accuracy, efficiency, and reproducibility, process automation including machine learning and artificial intelligence (AI) algorithms will likely be required.

So far, only the ion-implantation approach embeds qubits in isotopically enriched $^{28}$Si, which is achieved by high-fluence $^{28}$Si ion-implantation of a surface layer. Efforts are currently underway also in the STM-assisted approach to achieve epitaxial growth of $^{28}$Si in UHV. Yet, even in $^{28}$Si, electron and nuclear spin coherence times remain significantly shorter (of the order of 3 magnitudes) in nanoelectronic devices, compared to bulk ensembles (Table 1). The stark difference between the individual and bulk coherence times may be attributed to several factors. The presence of surfaces/interfaces in devices may lead to increased charge noise, although placing donor qubits within epitaxial single-crystal silicon has been shown to reduce noise levels to the shot noise limit [10]. Vicinal donors may allow dipolar flip-flop or possibly even hyperfine coupling of nuclear spins. Also, lattice strain and electric fields are known to affect spin life- and coherence times. Especially the latter, inherent to the gate-control of nanoelectronic devices, may lead to an enhancement of the spin-orbit coupling strength [11].

*Table 1: Summary of spin life- ($T_1$) and coherence times ($T_2$) measured in donor qubits in nanoelectronic devices (at T ≈ 100 mK), compared to bulk ensembles. $T_{2DD}$ denotes $T_2$ enhanced by dynamical decoupling. The measurement temperatures for the bulk ensembles range between 1-3 K.*

| Qubit type | Bulk ensembles | Nanoelectronic devices |
|---|---|---|
| Electron spin ($T_1$) | > hours [13-15] | > seconds [5] |
| Electron spin ($T_2$) | 10 s [14] ($^{28}$Si) | $T_2^*$ = 268 μs ($T_{2DD}$ = 0.56 s, $^{28}$Si) [8] |
| Nuclear spin ($T_1$) | > hours [13,15] | > days [5] |
| Nuclear spin ($T_2$) | $T_2$ = 27 s ($T_{2DD}$ = 3 h, $^{28}$Si) [15] | $T_2^*$ = 600 ms ($T_{2DD}$ = 35.6 s, $^{28}$Si) [8] |

Advances in Science and Technology to Meet Challenges

Even when single donors are placed with atomic level accuracy, for scalable architecture each qubit in the array needs to be addressed individually by local electric and magnetic fields without affecting their neighbors. The combined factors of the atomic-scale footprint of the qubits, the short-range nature of the exchange interaction, and the high gate-density it dictates, have all been contributors to the challenge of scaling to more than two donors [3,4,7,17-19]. To overcome scalability constraints, long-range qubit coupling schemes, such as the electrical dipolar interactions, are currently being explored [17,18] (Figs. 2A-C). An electric dipole may be formed by biasing a donor wave function close to the ionization point, hence causing a charge separation between the electron and the positive donor nucleus (Fig. 2B) [18]. Alternatively, combining a single donor with a donor molecule creates a finite hyperfine interaction between the two (Fig. 2A) [17] which may then be modulated by timedependent electric fields to induce electric dipole spin resonance (EDSR) of the electron and nuclear spins. Within this scheme, quantum information may be encoded in the states {| ↑⇓⟩ , | ↓⇑⟩}, for the single donor [18] and {| ⇓⇑⇑↓⟩ , | ⇓⇑⇓↑⟩} for the single donor-donor molecule

system [17] – forming so-called "flip-flop" qubits. Electric dipole coupling of up to tens of MHz and distances of hundreds of nm have been predicted [17,18], making this approach a viable long-range alternative in addition to exchange-coupled qubits. Qubit coupling may further be extended to mesoscopic or even macroscopic distances (~1 µm-1 cm) by coupling of spin qubits to microwave resonators [17,18] (Fig. 2C), possibly opening avenues for hybrid quantum technologies including both spin qubits and superconducting qubits (Fig. 2D). To further meet the challenges of scale-up, quantum error correction protocols, including topological surface codes, have been proposed [19]. Surface codes may be implemented within 2-dimensional qubit arrays controlled by a three-dimensional (3D) network of electrodes (Fig. 2E).

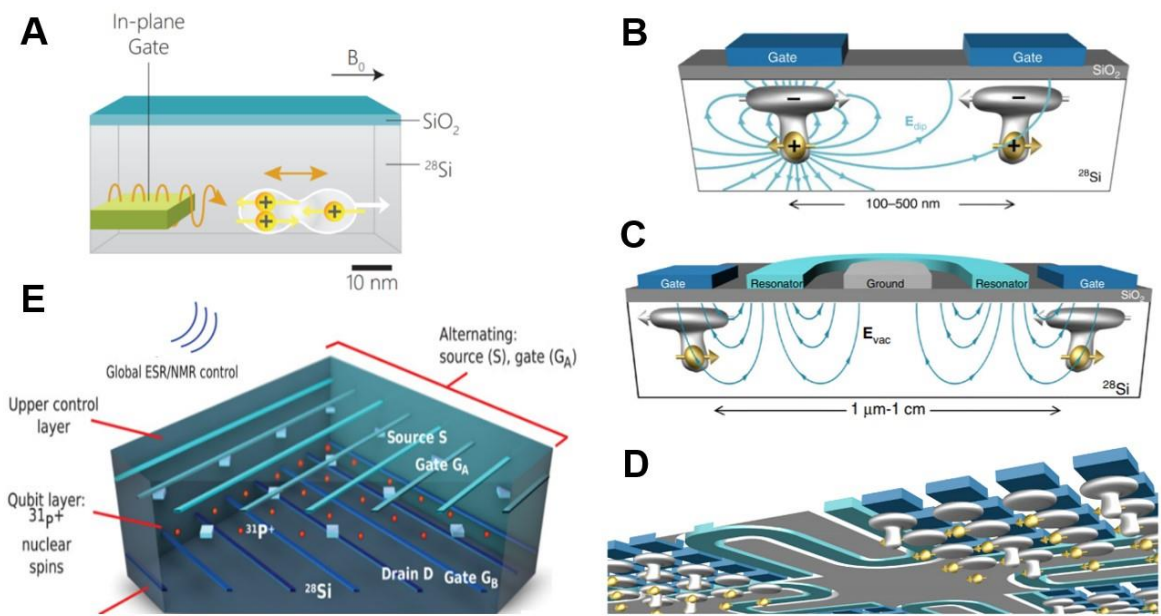

**Figure 2.** Long-range qubit coupling schemes and scale-up. (A) Schematic of a two-donor (1P-2P) qubit device. A charge dipole forms due to the asymmetric charge distribution which interacts with an in-plane gate electric field. The hyperfine coupling of this electric dipole can facilitate EDSR under the influence of an alternating electric field [17]. (B) Electric dipole-dipole interactions between two distant qubits [18]. Further scaling can be achieved by (C) coupling the electric dipole moment of donor qubits to microwave resonators, which extends the inter-qubit coupling distance. (D) A multi-qubit processor can then be built by coupling clusters of donor qubits [18]. (E) A schematic of multi-qubit quantum processor realizing a vertically stacked quantum error correction surface code architecture [19].

## Concluding Remarks

Silicon quantum devices, based on $^{31}$P qubits, rapidly maturing quantum technologies, with great promise should the extremely long lifetimes of electron and nuclear spins be harnessed in a scalable fashion. The two established fabrication techniques, i.e., ion-implantation and STM-assisted lithography, have so far enabled impressive demonstrations of rapid one-, two-, and three-qubit logic gates with high fidelity. Yet, scaling to several qubits and eventually a large-scale quantum processor remains a challenge due to the requirements of precisely controlled, efficient, and reproducible dopant placement in a low-noise qubit environment. Long-range qubit coupling schemes such as the electric dipole-dipole interaction may alleviate the requirement for precise donor placement but are yet to be experimentally realized. If proven viable, further coupling of Si:P qubits to microwave resonators may however present a path towards the exciting prospect of hybrid quantum technologies.

## Acknowledgements

This research was supported under the Singapore National Research Foundation (NRF) Fellowship (NRF-NRFF2017-11) "Atomic Manipulations in 2D Materials for Quantum Computing Applications" with further support from the Singapore Ministry of Education (MOE) Academic Research Fund Tier 3 grant (MOE2018-T3-1-002). BW and KEJG are further supported by the National Singapore Research

Foundation (NRF) Competitive Research Programme NRF-CRP21-2018-0001 and KEJG acknowledges the support of A*STAR DELTA-Q C210917001.

# 6. Color centers in wide-bandgap materials for quantum applications

Zhao Mu[1], Christoph Becher[2], Weibo Gao[1,]

[1]Division of Physics and Applied Physics, School of Physical and Mathematical Sciences, Nanyang Technological University, 637371, Singapore

[2]Fachrichtung Physik, Universität des Saarlandes, Campus E2.6, 66123 Saarbrücken, Germany

[3]The Photonics Institute and Centre for Disruptive Photonic Technologies, Nanyang Technological University, 637371, Singapore

**Status**

Color centers in semiconductor materials have been intensively studied since the first report of single negatively charged nitrogen vacancy (NV-) color centers in diamond. During the past two decades, their family members are growing from diamond NV- centers, to group IV split vacancy centers in diamond, silicon vacancy (SiV-) and divacancy (VV$^0$) centers in silicon carbide (SiC), G-centers in silicon, and the emerging $V_B^-$ centers in hexagonal boron nitride (hBN) [1]. Most of these defects have well-known molecule configuration and clear energy structure thanks to various experimental characterization methods and DFT calculations. These defects have distinct spin, optical, charge properties and additional functionality provided by their host materials along with nanofabrication. Exploiting these properties, they have enabled quantum sensing, quantum computation, and quantum communication.

Among the color centers pool, the diamond NV- center is the most studied system for quantum sensing, and quantum network applications [2]. The diamond NV- spin has a long coherence time and is a good sensor to the magnetic field, electrical field, temperature, and pressure. The DC and AC magnetic field sensitivity has been pushed to $pT/\sqrt{Hz}$ range [3]. The nanoscale NV- quantum sensor leads to many new applications in biology and condensed matter physics [4]. The long spin coherence and the rich nuclear resources surrounding NV- centers have enabled a quantum processor of 10 qubits with coherence time up to minutes [5]. Quantum error correction and fault-tolerant operation have also been realized, promising future quantum computation. A quantum network composed of three nodes [6] was demonstrated after the realization of the deterministic entanglement of two nodes.

Two major issues for diamond NV- centers are their low zero-phonon line (ZPL) ratio and their emission wavelength being sensitive to an external electric field, making it challenging to scale up. The group IV split vacancy centers with inversion symmetry, reducing their susceptibility to external fields, and a high ZPL ratio could be a potential candidate to realize quantum networks [7]. Recently, a two-qubit register (a SiV- electron spin and a nearby [13]C nuclear spin) was constructed, coupled efficiently to a diamond nanocavity. This efficient nanophotonic interface was further utilized for memory-enhanced quantum communications, a cornerstone of long-distance key distribution[8]. Beyond this, heterogeneous on-chip integration of diamond array containing group IV vacancies into an aluminum nitride photonic integrated circuit has been realized which may promise large-scale quantum information processing [9].

Apart from the diamond emitters, emitters in SiC benefit from the host material which could be grown in large scale, easy-doped, and nanofabricated into high-quality factor nanostructures. Taking these advantages, the emission wavelength of VV$^0$ centers can be tuned over Terahertz (THz) ranges and the emission linewidth can reach almost the lifetime limit when using charge depletion. The VV$^0$ defects also exhibit excellent spin properties of 5 seconds coherence time and can realize single-shot readout [10]. This progress demonstrates the SiC defects as a promising platform for realizing quantum network or quantum computation.

**Current and Future Challenges**

A common problem for both diamond emitters and SiC emitters for practical applications is the low photon extraction efficiency due to the total internal reflection. This deficiency will limit both the entanglement rates and the entanglement distribution distance, as well as the sensitivity of quantum sensors. The current solution to relax this limitation is to integrate color centers into nanostructured hosts[11], e.g. nanobeam cavities. However, these nanostructures often adversely impact the optical properties of color centers, i.e. their emission linewidths might be broadened due to spectral diffusion created by surface defects introduced during nanofabrication. Moreover, different emitters mostly exhibit different emission wavelengths due to different in-situ environment. This broadened emission and environment dependent emission wavelength created a barrier that must be overcome toward a fully chip-integrated quantum internet.

For building up a fiber-linked quantum network, both diamond $NV^-$ and group IV split vacancy have their challenges besides their visible emission wavelength. The state-of-the-art demonstration of multimode quantum networks is achieved with diamond $NV^-$ centers. The $NV^-$ centers have a small Debye-Waller factor (3%) and an environment-sensitive emission, suggesting a limited photon emission rate and imposing a limited qubit operation speed. The $SiV^-$ in diamond has a ZPL ratio of 70% which is beneficial for coherent photon generation. Nevertheless, the ground state spin coherence time is limited by the phonon bath which couples the ground state orbital doublet leading to the requirement of operating temperatures below 100 mK [12].

As a key player for quantum sensing, the main challenge for diamond $NV^-$ centers is the shortened coherence time and unstable charge state due to the fluctuation charges when the $NV^-$ centers are placed near to the surface. This makes it challenging for sensing of biological samples which requires nanodiamond hosts or for weak signal detection which needs a small, typically nanometer scale, sensor-target distance. On the other hand, SiC emitters seem to be potential sensor candidates given that both $SiV^-$ and $VV^0$ centers exhibit long coherence time and high ODMR contrast and the SiC host material is also biocompatible. However, the generation of these defects in SiC nanoparticles has rarely been studied. The $V_B^-$ centers in hBN are good sensors in terms of their high ODMR contrast and their 2D nature, but their short coherence time would limit their applications into AC field sensing.

**Advances in Science and Technology to Meet Challenges**

Optical transitions of $SiV^-$ emitters in SiC and $SiV^-$ emitters in diamond are predicted to be insensitive to the electrical field to the first order. The broadened emission of diamond and SiC emitters embedded into nanostructures are not intrinsic to themselves but due to the imperfect engineering process. A lifetime-limited emission linewidth is then expected by optimizing the fabrication recipe or exploiting the charge depletion method. Recently, $SnV^-$ centers in diamond nanostructures (nanopillars and waveguide)[13] and $SiV^-$ centers in reactive ion etching developed SiC waveguide were shown to exhibit lifetime limited emission [14]. Exploiting the PIN SiC junction, the charges surrounding the $VV^0$ centers are depleted, resulting in 50 times narrower emission linewidth[15]. Though these techniques or special treatments may alleviate or resolve the spectral diffusion problem, the once and for all solution still relies on the understanding of the behind mechanism.

Once the spectral diffusion problem is under control, the color center emission into the ZPL and the overall emission rates could be enhanced by overlapping the cavity resonance wavelength and the color center emission wavelength, as demonstrated for all the above-mentioned emitters in diamond and SiC. Regarding a single quantum node, the spin-photon entanglement rates could be increased by orders due to the Purcell enhancement of spontaneous emission and the potentially higher repetition rates. With regard to different nodes, the spin-spin entanglement requires indistinguishable photons emitted from separated nodes. Besides improvement in defect center generation technologies, in particular high-temperature annealing to reduce inhomogeneous broadenings, this requires an additional degree of freedom for tuning the emission wavelength which is provided by strain engineering for group IV split vacancy centers, and stark shift for color centers without inversion symmetry. By overlapping the cavities resonance frequency and color centers frequency, both on-chip quantum processors of several electron qubits and entanglement among multi-nodes could be conceived.

To bring the group IV split vacancy centers into cryogenic working temperature, the key is to decouple the ground orbital states from the phonons by augmenting the energy splitting between the ground doublet states. The strain engineering has allowed the ground state splitting to increase by 10 times for $SiV^-$ centers and prolonged the spin coherence time by a factor of 6 [16]. However, this benefit comes with the cost of breaking down the inversion symmetry. Alternatively, the $SnV^-$ and $PbV^-$ in diamond with greater splitting are potential choices. The $SnV^-$ exhibits a spin coherence time of 0.3 ms at 1.7 K [17], about 5 times longer than the $SiV^-$ coherence time at 100 mK.

During the last two years, several kinds of emitters in hBN are found to have spin sensors with prominent ODMR contrast. hBN has also been widely used to encapsulate 2D materials. The target-sensor distance thus could be brought within the sub nanometers range. Naturally, the 2D sensor could play a complementary role for specific applications like sensing of the magnetic field generated by 2D magnets. Regarding the diamond $NV^-$ centers, several approaches were used to prolong the coherence like surface engineering, dynamical decoupling, clock transition, coherent driving of surface electrons. However, these techniques still could not push the coherence to the $T_1$ limited level. A comprehensive study of the decoherence sources with a complete toolbox to take care of them is still needed.

The rich experience gained from studying the SiC and diamond color centers, along with the ongoing data mining assisted defect prediction would benefit the discovery of more defects which may outperform the current defect platforms or would be more suitable for certain applications [18]. These valuable resources would accelerate the study of emerging defects like hBN color centers and benefit the optimization of the optical and spin properties of the defect.

**Concluding Remarks**

The current color center platforms seem to be not yet perfect. It is still hard to predict whether one platform will deliver a solution for all applications in the end or whether different defects will turn out to be suited best for specific tasks. The class of color centers host is expanding from 3D materials to 2D materials, from wide-bandgap host to small-bandgap host. New color center discovery is also expected to be accelerated with the help of data-mining based prediction. With the increasing experience in the study of diamond and SiC color centers, both optical and spin properties of the existing platforms are improving with the combined efforts of materials engineering, enriched quantum control toolbox, matured nanofabrication techniques. This deep understanding of the pros and cons of the current platforms, in turn, benefits the goal of designing quantum defects for specific applications.

To realize a quantum network, integrating multiple color centers into nanostructures is the key ingredient but has not been fully demonstrated. The past progress was mainly realized with $SiV^-$ in diamond though the diamond nanofabrication is less matured compared to the SiC and Si counterparts. The current progress in $VV^0$ centers in SiC is attractive. The recent demonstration of the prolonged spin coherence via clock transition, enhanced emission into the ZPL, tunable emission wavelength, and single-shot readout of spin states would pave the way to spin-photon entanglement and even on-chip quantum processor benefiting from the matured SiC nanofabrication techniques.

The diamond NV$^-$ spin sensor has gained many successes in quantum sensing and quantum communication, mainly based on its long spin coherence. While it is still the leading platform for sensing applications, other systems with particular properties may play complementary roles. The VV$^0$ centers with clock transition may be favorable for Ramsey-type measurements while the hBN defects may be preferred for nanoscale sensing.

**Acknowledgements**

*We acknowledge Singapore National Research foundation through QEP grant (NRF2021-QEP2-01-P02, NRF2021-QEP2-03-P01, 2019-0643 (QEP-P2) and 2019-1321 (QEP-P3)) and Singapore Ministry of Education (MOE2016-T3-1-006 (S)).*

# 7. 2D Materials for Quantum Technologies

Jeng-Yuan Tsai and Qimin Yan
Department of Physics, Temple University, Philadelphia, PA 19122, USA

**Status**
Being atomically thin, amendable by defects and interfaces, and sensitive to external stimuli, two-dimensional (2D) materials offer a broad range of materials properties to serve as functional device components for a set of emerging quantum technologies, including quantum computing, quantum communication, and quantum sensing (Figure 1).

The development of solid-state qubit systems for quantum computation is crucial to the anticipated quantum revolution. 2D materials, including graphene, h-BN, and transition metal dichalcogenides (TMDs), have been explored as host materials in four qubit design strategies including quantum dots (QDs), defect spins, superconducting junction, and topological qubits[1].
Point defects in their hosts may create coherent quantum states for quantum computing and communication. The planar structures of 2D materials present a superior platform for realizing controlled creation and manipulation of defect qubits for scalability. Several quantum defects in h-BN with triplet spin states, including carbon-vacancy complex and negatively charged boron vacancy[2], has been proposed as spin qubits. Antisite defects in TMDs have been identified as viable spin qubits[3].

Using the spin states of trapped charges and the Coulomb blockade phenomenon, QD qubits can be created[4]. Coulomb-blockade behaviour was observed in graphene QD qubits and coupled multi-QD qubits have been proposed based on graphene nanoribbons. Owing to spin-orbit interaction and unique valley physics, qubits based on TMD QDs, including spin qubits, valley qubits and hybrid spin–valley qubits, have been theoretically explored[5]. Coulomb blockade in encapsulated $MoS_2$ QDs[1] and $WSe_2$ QDs[6] has been observed. Solid-state qubits can also be realized using superconducting circuits based on Josephson junctions (JJs). The diverse properties of 2D materials provide the opportunity to fabricate all-2D JJ, eliminating device complexities.
Optical transitions created by deep defect states or defect-bound excitons in single-photon emitters (SPEs) can be utilized for quantum communication technique. In 2015, the first SPEs in 2D materials are demonstrated in monolayer $WSe_2$, a semiconductor compound in the TMD family[7, 8]. Later on, h-BN emerged as a promising material platform of polarized high-brightness SPEs that operate at room temperature[9]. The emission source in h-BN have been identified to be carbon related. Going beyond individual 2D materials, quantum emitters have been observed in twisted 2D heterojunctions from moiré-trapped excitons[10].

As the key component of quantum sensing technology, solid-state quantum sensors use atom-like systems, including defects in solids, with discrete and tunable energy levels to measure perturbations in the environment. While providing natural protection, the use of bulk materials limits the sensitivity and impedes on-chip integration. 2D materials with defects can greatly improve the versatility and sensitivity of quantum-sensing technologies. Recently, negatively charged boron vacancies in h-BN have been identified as promising quantum sensors for temperature, pressure, and magnetic field [11].

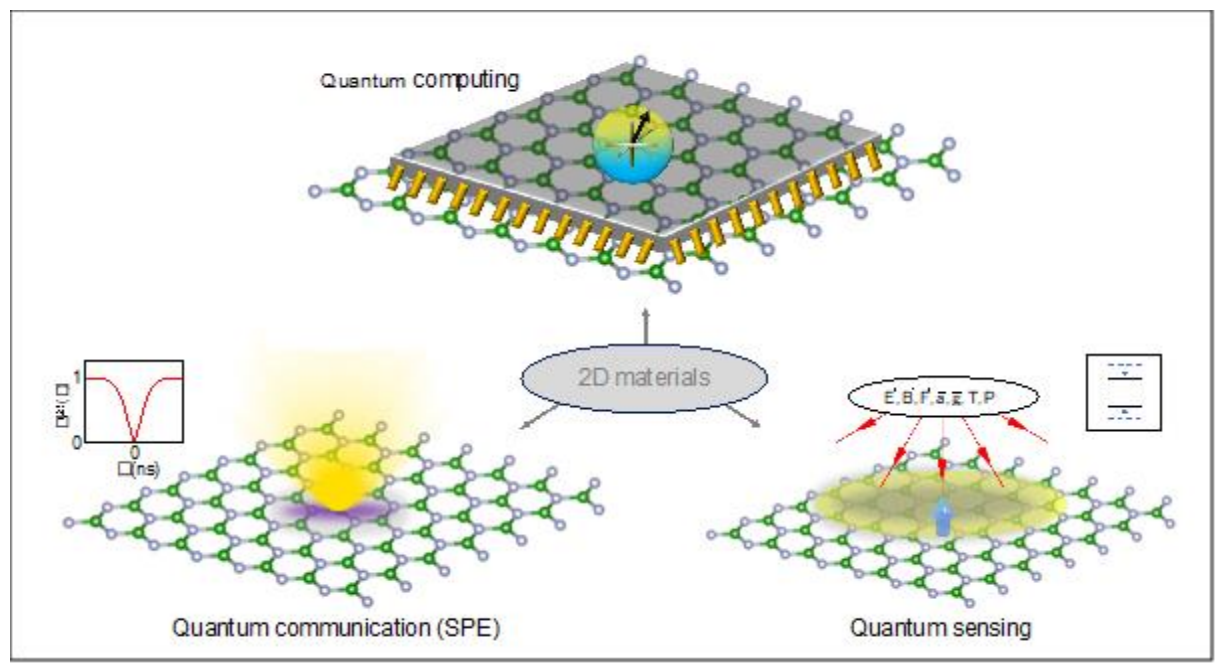

Figure 1. Applications of 2D materials for emerging quantum technologies, including quantum computing, quantum communication, and quantum sensing.

**Current and Future Challenges**

Despite its vast potential, the field is still in its early stage and full exploitation of 2D materials for quantum technologies are facing challenges that call for advances in science and technologies.

In the field of qubits in 2D materials, great challenges remain for qubit fabrication, characterization, and device integration. It is still a great challenge to create stable defect qubits in h-BN or TMDs in a controllable manner, which is important for deterministic qubit fabrication and scalability in the long run. Another challenge is to provide the protection for defect qubits against environmental perturbation. For QD qubits, the imperfect edges of graphene create disorders that compromises quality and reproducibility even with h-BN encapsulation. The proposed spin and valley qubits in semiconducting TMD QDs are still waiting for experimental realization. Going forward, major long-term challenges are scaling up qubit circuits and realizing quantum computation based on coordinated quantum networks.

For SPEs in 2D materials, there are challenges to improve the performance of these emitters in terms of indistinguishability, room-temperature accessibility, electrical and photonic addressability, and entanglement. For instance, entangled light sources have not been demonstrated in any 2D materials yet. It is challenging to unravel the origins of the defects that are responsible for the observed quantum emission in h-BN. Also, it is necessary to identify SPEs in other 2D materials that cover a broader spectrum (from near infrared to visible) for usage in quantum technologies such as optical-fiber-based quantum communication.

In the field of quantum sensing, the demonstration of boron vacancy quantum sensors in multi-layer h-BN[11] is an encouraging starting point, but there is still a long way to go before it reaches the ultra-thin monolayer limit. The 2D nature of monolayer van der Waals materials is expected to be a double-edged sword for quantum sensing. With substantial charge or spin noise from the environment, there is still an unanswered question that whether quantum defects in a specific charge state can exist in monolayers and what the sensing performance would be in that case. A long-term challenge is to identify quantum defects in other 2D materials in the single defect limit for nanoscale sensing.

In general, full exploitation of 2D materials for quantum technologies faces challenges in fabrication methods and large high-quality materials are pivotal to practical deployment of the 2D quantum platform. Another general challenge in the field is to go beyond the existing 2D materials.

**Advances in Science and Technology to Meet Challenges**

Moving forward, unprecedented advances in 2D materials growth and characterization have been achieved, which will meet the major challenges in deploying 2D materials for quantum technologies.

Synthetic strategies and assembly schemes with wafer-scale homogeneity have been developed for 2D materials such as h-BN and TMDs. With the advance of material fabrication techniques, 2D materials can be interfaced with other materials to form multifunctional heterostructures, opening new possibilities for on-chip quantum communication and quantum sensing.

Defect fabrication in 2D materials can be improved by using low-energy ion implantation and bottom-up synthesis methods based on the achievement made in diamond. The development of scanning tunnelling and transmission electron microscopy provide the capability to image individual defects in 2D solids with atomic resolution. Rapid progress is being made in physical characterization of SPEs, such as super-resolution imaging and *ab initio* modelling of defect-based emission sources. The development of techniques to integrate SPEs in 2D quantum hosts with atomically thin lenses and metasurfaces for light manipulation will lead to integrated quantum nanophotonic network and other quantum devices.

*ab initio* computational methods associated with defect theory is being actively developed toward fundamental understanding of various aspects of quantum defect states[2, 12], such as excitations and relaxations, spin coherence, correlation effects, and intersystem crossing phenomena. Meanwhile, the continued development of various QD systems based on TMDs, superconducting 2D materials, and their heterojunctions will create multiple alternative and promising platforms for realizing diverse types of qubits for quantum technologies.

Another major advance associated with 2D materials is the fabrication of 2D heterostructures with functionalities unavailable in bulk materials. The continued development of encapsulation/insulation techniques by fabricating 2D heterojunctions will provide qubits and SPEs with protection against environmental effects. The understanding in strongly correlated physics and topology in quantum materials has enabled 2D heterojunction systems as a robust quantum simulation platform[13]. With the recent development of synthetic 2D materials, the number of building blocks for heterojunction-based quantum systems will continue to grow.

Going beyond h-BN and TMDs, the development of data-driven materials discovery approaches such as the construction of 2D materials databases with defect information[14] will accelerate the search for other 2D material systems that enable integration in solid-state, on-chip quantum devices. With the aid of data-driven approaches[3, 14], the discovery of stable and coherent qubits and SPEs in novel 2D materials systems can bring a breakthrough to emerging quantum technologies, even at room temperature.

**Concluding Remarks**

2D materials host inherent advantages toward the development of emerging quantum technologies, including quantum computing, quantum communication, and quantum sensing. The emergence of a diverse set of quantum phenomena in 2D materials offer great potential for developing integrated quantum technologies. In addition, heterostructures of 2D materials offer a new quantum design platform with functionalities unavailable in individual materials. Despite significant advances in 2D material synthesis and characterization, 2D quantum technologies are far from mature and major challenges remain. With the rapid progress in scientific and technological developments in the field, the future of 2D materials for quantum technologies is bright.

**Acknowledgements**

This work was supported by the U.S. Department of Energy, Office of Science, Basic Energy Sciences, under Award #DE-SC0019275.

## 8. Superconducting Materials for Single-Photon Detectors
Samuel Gyger, Stephan Steinhauer, Val Zwiller
KTH Royal Institute of Technology, Stockholm 106 91, Sweden

**Status**
Quantum information processing has grown to extraordinary prominence within today's research and development environments, with superconducting materials playing a decisive role for the related device technologies. Immense progress in different fields has relied on continuous advances in our knowledge on superconducting structures at mesoscopic and nanoscopic scales, including their interactions with electromagnetic radiation from microwave to optical frequencies. To date, superconducting materials are key enablers for a variety of emerging quantum technologies with the potential for significant societal and commercial impact, ranging from superconducting quantum computing to single-photon detectors.

Research efforts over the last several decades have established that superconducting devices can offer unrivalled capabilities for the detection of single photons in the visible and near-infrared wavelength range. Based on the photon absorption-induced effects in superconducting materials, different operation principles have been elaborated, most notably kinetic inductance detectors, transition edge sensors and superconducting nanowire single-photon detectors (SNSPDs). The latter technology, initially demonstrated in 2001 [GOL01], provides exciting prospects for a wide range of applications in modern photonic quantum technologies due to the unique combination of close-to-unity detection efficiencies exceeding 98%, high time resolution below 10ps, suitability for operation in compact cryocoolers and comparatively simple readout schemes [HOL19, ZAD21]. In this roadmap perspective, we will hence focus on SNSPD technology and related materials science aspects as well as challenges, providing an outlook on future directions. Note that, recently, microwire-based superconducting devices exhibiting single-photon detection capabilities have been obtained, relying on comparable operation principles.

Previous advances in single-photon detectors based on SNSPDs are summarized in Figure 1, presenting a timeline of technological milestones. The relevant superconducting materials and their first use for detector fabrication have been compiled, together with selected performance improvements in terms of system detection efficiency (SDE) at telecom wavelengths (1550nm) and timing resolution / jitter. While the significant evolution in the field has also been based on progress in nanofabrication, photonic packaging, measurement technology and cryogenics, the growth of optimized superconducting materials with tailored properties has certainly played a vital role in these developments which cannot be underestimated.

**Current and Future Challenges**
The fabrication of SNSPD devices relies on the deposition of superconducting thin films with thicknesses around 10 nm and below. The growth process is linked with considerable requirements on homogeneity and process control, as the device properties and performance crucially rest on various, often inter-dependent, material parameters, such as superconducting gap, scattering rates for electrons and phonons, crystalline structure, level of disorder / defects, thickness and morphology. During the last two decades, magnetron sputtering has served as workhorse for the fabrication of superconducting thin films employed for SNSPD devices; if not mentioned otherwise, the prior reports reviewed here have relied on this deposition technique.

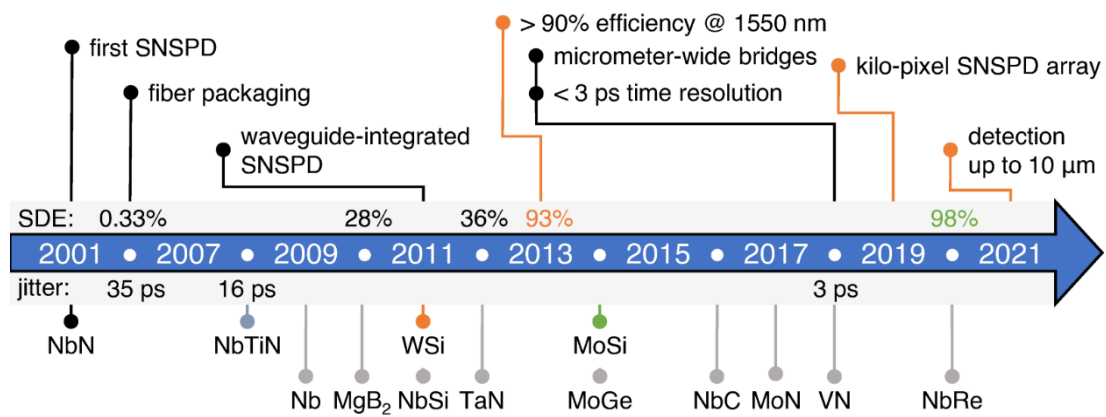

Figure 1. Timeline of key developments in the field of superconducting nanowire single-photon detectors (SNSPDs) from their initial demonstration in 2001 until to date. Furthermore, advances in the performance parameters system detection efficiency (SDE) at telecom wavelengths (1550nm) and timing resolution / jitter are shown, together with the superconducting materials employed for detector fabrication (note the colour coding).

The substantial evolution in SNSPD device performance and functionality has progressed along several fronts: The detector SDE has almost reached its fundamental limit of unity efficiency, while further optimization will most probably rely on improving optical coupling and light absorption in the detector structure, at least in the visible and near-infrared wavelength range. For the detection of long-wavelength photons in the infrared, superconductors with a small superconducting gap and/or reduced free carrier density are beneficial [VER21] – a challenge that can be addressed by the choice of material and stoichiometry optimization. On the other hand, the physical limitations for achieving highest timing resolution / lowest jitter have remained elusive, with the current record value reported being below 3ps. While multiple factors contribute to the overall detector jitter (e.g. electronic noise, signal propagation delay), it has been proposed that inhomogeneity of the superconducting material as well as its inelastic scattering time contribute to the intrinsic jitter [ALM19]. In this regard, engineering superconducting device structures that minimize intrinsic jitter constitutes a major challenge for realizing SNSPDs with optimized timing resolution. To further boost the functionality of SNSPD-based detector technology, it will be required to massively increase the active area and the number of pixels, for instance for single-photon imaging and large-scale photonic integrated circuits. Recent progress [STE21] has resulted in arrays with more than 1000 pixels and active areas approaching 1mm$^2$, which still significantly falls short of other single-photon detector technologies. From a materials science point of view, the major challenge connected to SNSPD up-scaling is the limited homogeneity of the superconducting thin films employed for device fabrication. Moreover, to address challenges related to the cryogenic cooling system, superconducting materials suitable for detector operation at higher temperatures (for current devices most commonly below 3K) would be particularly valuable.

**Advances in Science and Technology to Meet Challenges**
Future innovation will critically depend on the availability of specifically optimized superconducting materials, obtained either via refining widely-used magnetron sputtering processes further or via establishing alternative growth techniques. Epitaxial NbN thin films (thickness 7nm) on AlN-on-sapphire substrates have been achieved with molecular beam epitaxy (MBE), resulting in single-crystalline layers with the NbN {111} axis aligned with the AlN c-axis [CHE20]. Narrow nanowire linewidths down to 20nm were required to approach saturated internal detection efficiency in the near-infrared. Nevertheless, this report confirmed the potential of MBE-grown films, as they showed high critical temperatures as well as lower sheet resistance and considerably higher critical current density compared to their poly-crystalline counter-parts. Further research will be needed to elaborate how single-crystalline films with such a low degree of disorder can be optimally utilized for single-photon detectors and if they can provide advantages in terms of homogeneity for high-yield fabrication. Another avenue towards the latter is the use of plasma-enhanced atomic layer deposition (PEALD) for superconducting material growth. In general, ALD is considered a deposition method well-suited for achieving layers with excellent uniformity that are pinhole-free and conformal to the substrate, which is highly desirable from an SNSPD perspective. The viability of growing NbN thin films with PEALD has been shown, resulting in prospects for large-scale fabrication and for achieving detection capabilities further into the infrared [TAY21].

In addition, future advances could be a result of innovation linked with the use of alternative superconducting material systems. A notable example in this regard is $MgB_2$, which exhibits high critical temperatures above 30K and low kinetic inductance (30-60 times lower compared to NbN). High-quality $MgB_2$ thin films (thickness 5nm) have been achieved with hybrid physical-chemical vapour deposition and SNSPDs with only 130ps relaxation time have been obtained [CHE21], making such devices ideally suited for operation at high photon count rates. Other research efforts have been directed towards developing SNSPD devices with cuprate high-temperature superconductors, which are particularly appealing due to the high critical temperatures above the liquid nitrogen boiling point. To date, major challenges are still lying ahead, ranging from solving technological difficulties (e.g. material instability, patterning processes) to unveiling the physical mechanisms and operation conditions for single-photon detection. However, recent results on $YBa_2Cu_3O_{7-x}$ nanowires, demonstrating hotspot formation and phase slip control induced by single photons, hold great promise for the successful realization of devices with superior timing resolution in the future [LYA20].

**Concluding Remarks**
Advances in understanding and controlling superconducting material properties at the nanoscale have played a key role for achieving SNSPD devices with superior performance and their burgeoning commercialization. Recent years have shown significant improvements in thin film deposition technology as well as the emergence of various growth techniques and material systems. To guide future developments, it will be crucial to identify structure-property relationships of the relevant superconducting materials, correlating parameters such as crystalline structure, chemical composition and defect density with transport characteristics and SNSPD performances. Studying the employed materials and devices from a materials science point of view will not only enable further detector optimization, but also deepen our insights into the fundamental operation principles. This will give us the opportunity to reconcile successful detector concepts with promising material systems, specifically geared towards innovative solutions for quantum technology applications and beyond.

**Acknowledgements**
We acknowledge the support from the European Union's Horizon 2020 Research and Innovation Action under Grant Agreement No. 899824 (SURQUID) and No. 899580 (FastGhost).

[ALM19] J.P. Allmaras, A.G. Kozorezov, B.A. Korzh, K.K. Berggren, M.D. Shaw, "Intrinsic Timing Jitter and Latency in Superconducting Nanowire Single-photon Detectors," Phys. Rev. Appl., vol. 11, no. 3, 034062, 2019.

[VER21] V. B. Verma, B. Korzh, A. B. Walter, A. E. Lita, R. M. Briggs, M. Colangelo, Y. Zhai, E. E. Wollman, A. D. Beyer, J. P. Allmaras, H. Vora, D. Zhu, E. Schmidt, A. G. Kozorezov, K. K. Berggren, R. P. Mirin, S. W. Nam, M. D. Shaw, "Single-photon detection in the mid-infrared up to 10 μm wavelength using tungsten silicide superconducting nanowire detectors," APL Photonics, vol. 6, no. 5, 056101, 2021.

[STE21] S. Steinhauer, S. Gyger, V. Zwiller, "Progress on large-scale superconducting nanowire single-photon detectors," Appl. Phys. Lett., vol. 118, no. 10, 100501, 2021.

[CHE20] R. Cheng, J. Wright, H. G. Xing, D. Jena, H. X. Tang, "Epitaxial niobium nitride superconducting nanowire single-photon detectors," Appl. Phys. Lett., vol. 117, no. 13, 132601, 2020.

[TAY21] G. G. Taylor, D. V. Morozov, C. T. Lennon, P. S. Barry, C. Sheagren, R. H. Hadfield, "Infrared single-photon sensitivity in atomic layer deposited superconducting nanowires," Appl. Phys. Lett., vol. 118, no. 19, 191106, 2021.

[CHE21] S. Cherednichenko, N. Acharya, E. Novoselov, V. Drakinskiy, "Low kinetic inductance superconducting $MgB_2$ nanowires with a 130ps relaxation time for single-photon detection applications," Supercond. Sci. Technol., vol. 34, no. 4, 044001, 2021.

[LYA20] M. Lyatti, M.A. Wolff, I. Gundareva, M. Kruth, S. Ferrari, R.E. Dunin-Borkowski, C. Schuck, "Energy-level quantization and single-photon control of phase slips in $YBa_2Cu_3O_{7-x}$ nanowires," Nat. Comm., vol. 11, 763, 2020.